# Aluminium fast neutron leakage spectrum validation

Martin Schulc[a], Tomáš Czakoj[a], Evžen Novák[a], Alena Krechlerová[a], Adam Greš[a], Jan Šimon[a], Bohumil Jánský[a], Jiří Rejchrt[a], Michal Košťál[a], and Zdeněk Matěj[b]

[a] Research Centre Rez Ltd, 250 68 Husinec-Řež 130, Czech Republic
[b] Masaryk University, Botanická 68a, Brno 602 00, Czech Republic

Email: Martin.Schulc@cvrez.cz

Abstract

Aluminum is a crucial material in the nuclear industry, valued for its ability to perform reliably over time. The manuscript focuses on validating aluminum neutron transport libraries. The validation was conducted by activating samples in different positions, and measuring the fast neutron spectrum in the energy range of 0.1 MeV to 1.3 MeV by means of a proportional detector filled with hydrogen and, in the energy range of 1-12 MeV using a scintillation stilbene detector. Validation experiments were performed on the aluminium block with a central hole which was assembled from smaller aluminum plates. The dosimetric reactions studied for validation purposes were $^{58}$Ni(n,p)$^{58}$Co, $^{197}$Au(n,g)$^{198}$Au, $^{63}$Cu(n,g)$^{64}$Cu, $^{181}$Ta(n,g)$^{182}$Ta, $^{92}$Mo(n,p)$^{92m}$Nb, $^{nat}$Ti(n,X)$^{46}$Sc, and $^{nat}$Ti(n,X)$^{47}$Sc. All experimental data were compared to MCNP6.2 simulations using the ENDF/B-VIII.0, JEFF-3.3, and JENDL-5 neutron transport libraries. Activation cross sections were taken from IRDFF-II library. Concerning activation reactions results, unsatisfactory results are achived for $^{197}$Au(n,γ)$^{198}$Au reaction regardless of thickness and library. All other results are reasonable regardless on library. JEFF-3.3 and ENDF/B-VIII.0 fast neutron flux calculations are similar. All libraries give results within two sigma uncertainty except JENDL-5 in the region of 1.8-3.4 MeV, and ENDF/B-VIII.0 (JEFF-3.3) in the region of 1.06 – 1.3 MeV.

Introduction

Aluminum was chosen for the experimental study due to its widespread use as an important structural component in the nuclear energy industry. The results of these integral experiments, conducted in simple geometry like the slab configuration used here, can be easily compared with calculations based on different nuclear data libraries. Such experiments are invaluable for validating neutron cross-section data, as integral quantities such as neutron flux can often be measured more accurately than differential nuclear data. Therefore, these data provide a valuable opportunity to validate and/or refine nuclide cross-section evaluations. Series of measurements have been conducted in the past involving neutron flux passing through layers of various materials and geometries [1-2].

The neutron leakage and spatial distribution of neutron flux in an aluminum block were assessed with a $^{252}$Cf (s.f.) neutron source positioned at the block's geometric center. Various detectors were placed at different locations to conduct these measurements. The neutron leakage spectrum was measured using the proton recoil method at a distance of 72.8 cm from the surface of the block. For the energy range of 0.1 MeV to 1.3 MeV, a spectrometer with a proportional detector filled with hydrogen (HPD) was used, with the flux evaluated in 40 groups per decade. In the interval 0.9 MeV to 13.0 MeV, a stilbene crystal with dimensions ⌀ 10 mm × 10 mm.was used, and the flux was evaluated in 0.1 MeV groups. The applied methodology and techniques for measuring and evaluating the fast neutron flux in spherical benchmark arrangements are presented in [1,3] and are used in the aluminum block experiment as well.

In the second set of experiments, the spatial distribution of activation within the aluminum block was examined. The block was constructed from aluminum plates, each with a nominal thickness of 5 cm, with foils inserted between the plates to ensure their alignment along the axis of the neutron source. These foils, which were sensitive to thermal, epithermal, and fast neutrons, were used to measure their activity.

Experiment Description

The fast neutron leakage flux from an aluminum block (50.18 cm × 50.12 cm × 54.42 cm) and the spatial distribution of reaction rates measurements were conducted using a $^{252}$Cf (s.f.) neutron source located at the block's center. The experimental block was constructed from aluminum plates, each with side dimensions of 50.18 cm × 50.12 cm and a nominal thickness of 4.95 cm. Eleven plates were assembled to form a 54.42 cm thick block (see Figure 1). The plates were clamped together using carpenter's steel clamps and placed on an aluminum supporting table. The composition of the block is listed in Table 1. The neutron source was placed in a transport capsule at the end of a pneumatic transport system, ensuring that the source was positioned at the block's center during measurements.

*Table 1: Composition of the aluminium block according to XRF analysis.*

| Element | Mass fraction | Rel. unc. [%] |
|---|---|---|
| Al | 0.9973873 | 0.2 |
| Fe | 0.0023920 | 9 |
| V | 0.0001696 | 7 |
| Mn | 0.0000510 | 50 |

$^{252}$Cf (s.f.) neutron source placed in the drilled hole at the center of the block. The hole diameter was 2.81 cm. The total neutron emission from the $^{252}$Cf source varied from 2.07*10$^8$ n/s to 1.85*10$^8$ n/s during the measurements. The source emission was verified through a manganese sulfate bath at the National Physical Laboratory in the United Kingdom.

Seven types of activation detectors in foil form (AF) were used in the experiment, which was based on reaction rate measurements. These detectors included pure indium, gold, copper, titanium, nickel, tantalum, and molybdenum. The irradiation experiment was conducted in two separate phases. In the first phase, indium foils with a diameter of 3.5 cm and a thickness of 0.6 mm were placed in the central plane, shielded by aluminum of varying thickness. In the second phase, sets of foils were positioned in the gaps between different aluminum plates. Figure 2 illustrates the arrangement of the in the second phase of the experiment.

Two types of gold foils were used: one with a diameter of 1.5 cm and a thickness of 0.05 mm, which was placed in the position closest to the neutron source, and others with a diameter of 3 cm and a thickness of 0.06 mm, which were used in the remaining positions. Copper foils were cylindrical, with a diameter of 1.5 cm and a thickness of 0.254 mm. Nickel foils were square, measuring 1 cm by 1 cm with a thickness of 1 mm. Tantalum foils were also square, each measuring 0.8 cm by 0.8 cm with a thickness of 0.127 mm. Titanium foils were square, each measuring 1.6 cm by 1.6 cm with a thickness of 1 mm. Molybdenum foil was rectangular, measuring 1.55 cm by 1.52 cm with a thickness of 0.1 mm.

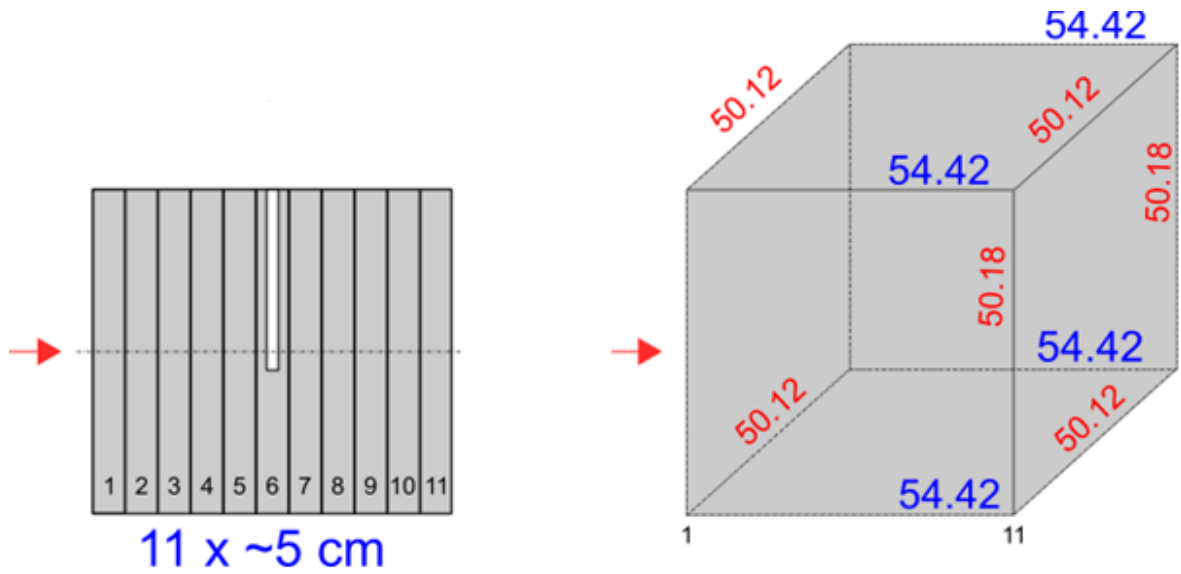

Figure 1: Scheme of aluminium block used in experiments, all dimensions are plotted in cm.

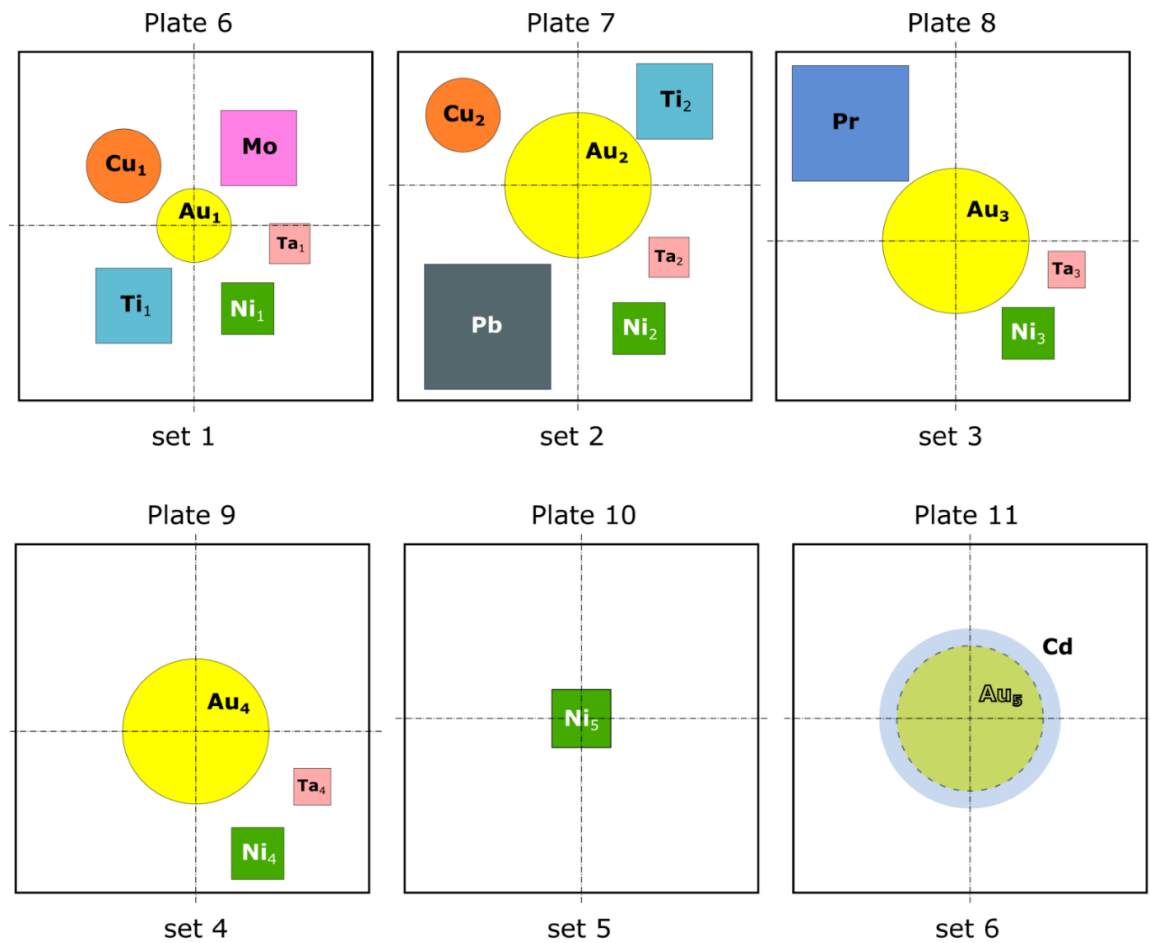

Figure 2: Illustration of activations foils positions relative to the position of $^{252}$Cf Source (centre), not to scale.

Measurements of reaction rates at various positions inside the aluminum block were conducted using dosimetry reactions. Firstly, pure indium foils were used to measure the $^{115}$In(n,n')$^{115m}$In reaction at different depths within the block. Secondly, set of foils was placed into inner block positions, including pure nickel for the $^{58}$Ni(n,p)$^{58}$Co reaction, pure gold for the $^{197}$Au(n,γ)$^{198}$Au reaction, pure copper for the $^{63}$Cu(n,γ)$^{64}$Cu reaction, pure tantalum for the $^{181}$Ta(n,γ)$^{182}$Ta reaction, pure titanium for the $^{nat}$Ti(n,x)$^{47}$Sc and $^{nat}$Ti(n,x)$^{46}$Sc reactions, and pure molybdenum for the $^{92}$Mo(n,p)$^{92m}$Nb reaction. All mentioned reactions have validated cross sections in the standard $^{252}$Cf neutron field, see [4].

Net Peak Areas for the activation products induced in the foils were measured using a well-characterized HPGe detector (ORTEC GM35P4) [5]. The detector has validated material and geometrical data which allowed for precise calibration of efficiency. The detector efficiency is calculated using a MCNP model. Activation foils were positioned on the detector's end cap during all measurements.

The measured NPA was used for the determination of the reaction rates, defined by Eq. **Chyba! Nenalezen zdroj odkazů.**:

$$q = \frac{C(T_m)\lambda T_m}{\eta \varepsilon N k T_l} \frac{1}{e^{-\lambda \Delta T}} \frac{1}{1-e^{-\lambda T_m}} \frac{1}{1-e^{-\lambda T_{irr}}}, \qquad (1)$$

where: $q$ is the experimental reaction rate per atom per second, $N$ is the number of target isotope nuclei, $\eta$ is the detector efficiency, $\varepsilon$ is the gamma branching ratio, $\lambda$ is the decay constant, $k$ characterizes the abundance of the isotope of interest in the target and its purity, $\Delta T$ is the time between the end of irradiation and the start of HPGe measurement, $C(T_m)$ is the measured number of counts per second, $T_m$ is the real time of measurement by HPGe, $T_l$ is the live time of measurement by HPGe (it is time of measurement corrected to the dead time of the detector), and $T_{irr}$ is the time of irradiation.

The resulting value is the relevant reaction activation rate $q$ for the specific material. This reaction rate is proportional to the neutron flux, accounting for its neutron energy spectrum. Using dosimetric reactions with low uncertainties is an excellent tool for monitoring neutron leakage. The radionuclides under study half-lives, evaluated gamma energy lines, and gamma emission probabilities are summarized in the Table 2.

*Table 2: Parameters of the investigated reactions.*

| Reaction | Half-life | Gamma Energy [MeV] | Gamma emission probability |
|---|---|---|---|
| $^{197}$Au(n,g)$^{198}$Au | 2.6941 days | 0.411802 | 95.62 % |
| $^{63}$Cu(n,g)$^{64}$Cu | 12.701 hours | 0.511 | 35.2 % |
| $^{181}$Ta(n,g)$^{182}$Ta | 114.74 days | 1.221395 | 27.23 % |
| $^{115}$In(n,n´)$^{115m1}$In | 4.486 hours | 0.336241 | 45.9 % |
| $^{58}$Ni(n,p)$^{58}$Co | 70.86 days | 0.81076 | 99.45 % |
| $^{92}$Mo(n,p)$^{92m}$Nb | 10.15 days | 0.93444 | 99.15 % |
| $^{nat}$Ti(n,X)$^{46}$Sc | 83.79 days | 0.889277 | 99.984 % |
| $^{nat}$Ti(n,X)$^{47}$Sc | 3.3492 days | 0.159381 | 68.3 % |

Neutron leakage spectrum from the block was carried out using the proton recoil method with digital spectrometers. Four measurements were conducted using both the stilbene and HPD detector. The methodology and techniques used for measuring and evaluating the fast neutron flux was described in [1,3].

The neutron flux in the energy range 0.1 MeV to 1.3 MeV was measured using a multichannel analyzer and a detection system with two spherical proportional detectors. The method of recoil protons measurement, derived from the elastic scattering of neutrons on hydrogen nuclei, was employed to determine the fast neutron spectra. The HPD detectors were filled with pure hydrogen gas at different pressures: 400 kPa, covering the energy interval from 0.1 MeV to 0.8 MeV, and 1000 kPa, covering the energy interval from 0.2 MeV to 1.3 MeV. The spherical shape of the detector was chosen for its incident neutron angle-independence.

The stilbene detector covering range of 0.9 MeV-13 MeV uses the proton recoil method with neutron and gamma pulse shape discrimination through digital processing of the detector signal. More about this two-parametric spectrometric system can be found in [6,7].

In an environment with a high gamma background, the separation between neutrons and gammas can be complicated, resulting in higher neutron detection thresholds, which is reflected in higher thresholds of neutron detection. In the studied case of aluminium leakage experiment, both neutron and gamma signals are well-evaluated even at a neutron energy of 0.86 MeV, see Figure 4 and Figure 5. The energy of the detected particle is the parameter determined by the spectrometer which is evaluated from the integral of the whole response to the incident particle.

Neutron spectra measurements were performed in two steps: first one with shielding cones and second one without shielding cones. The shielding cones are composed of iron and borated polyethylene to capture neutrons. The measurement with shielding cones served to estimate the background, which was then subtracted from the measurement without cones to obtain the signal. The dimensions and placement of the shielding were optimized to suppress all neutrons traveling directly from the block to the detector, see Figure 6.

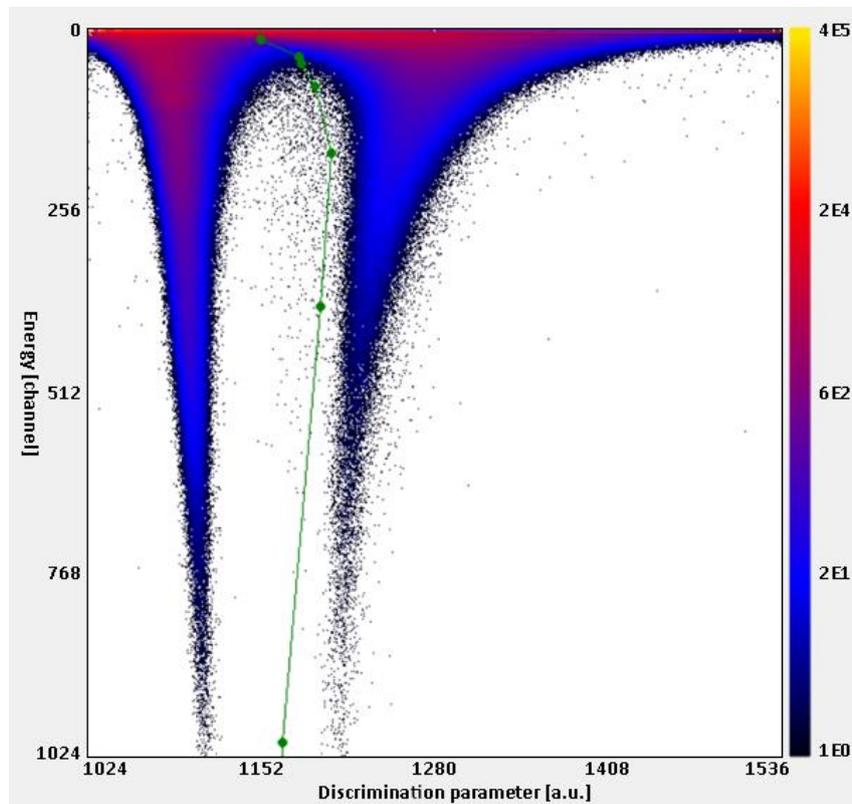

Figure 4: Measured two parametric leakage flux from aluminium block. The separation line is green, the gamma part (recoil electrons) is left, the neutron part (recoil protons) is right.

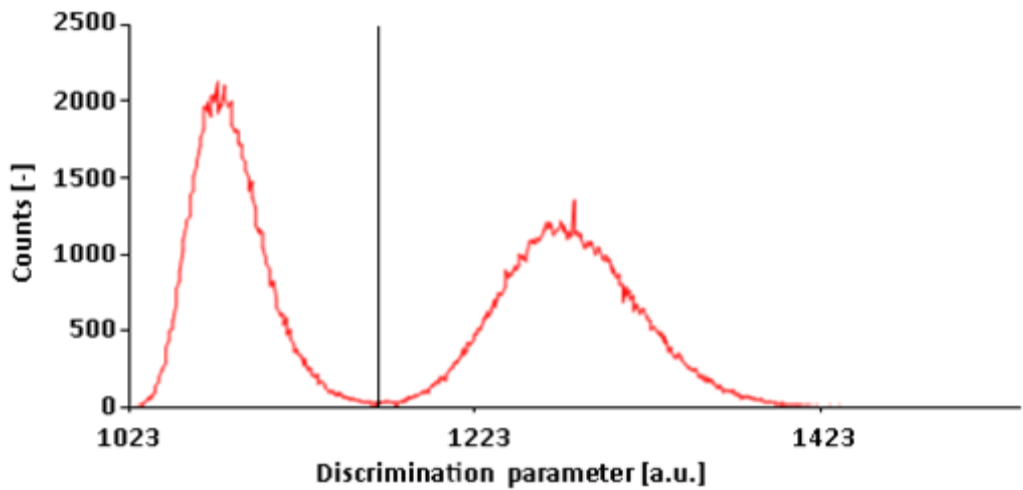

Figure 5: Discrimination between gammas (left) and neutrons (right).

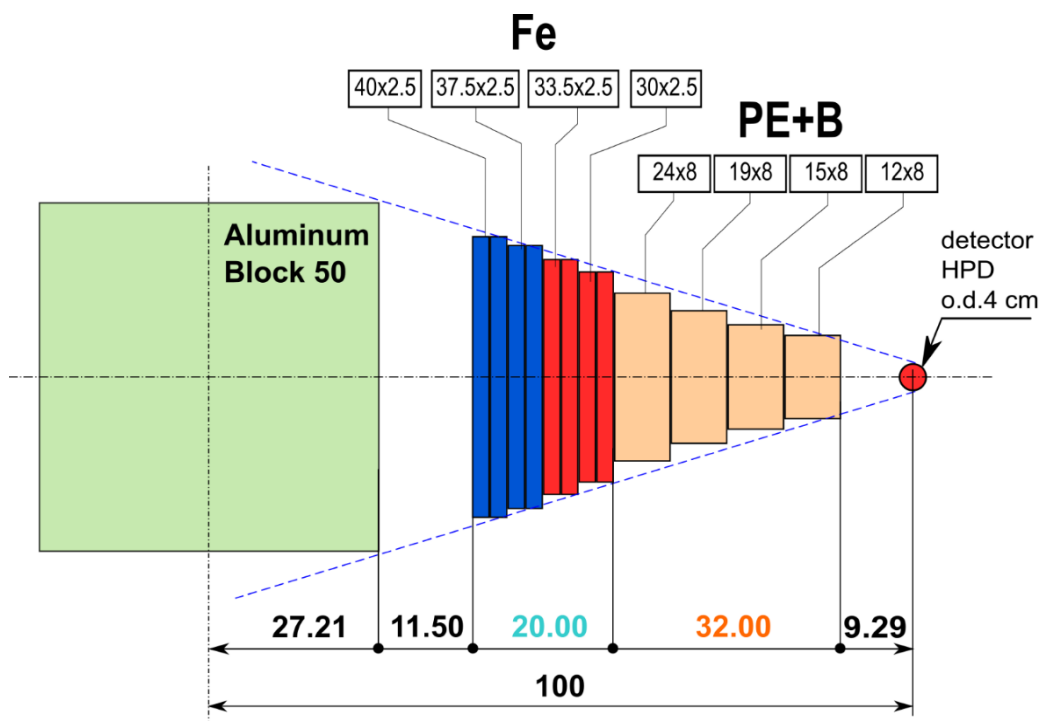

Figure 6: Aluminium block with shielding cones (iron and borated polyethylene cylinders) and HPD detector, dimensions in cm.

The relevant uncertainties which were taken into account were: uncertainty in the positions and dimensions of the samples and detector, aluminium density, total neutron emission of $^{252}$Cf the source, energy and efficiency calibration uncertainty of the detectors, and statistical uncertainty in the energy bins. The detailed uncertainty analyis of the measurements with $^{252}$Cf source were published in [8].

Monte Carlo transport simulations

The simulation of neutron transport was calculated using the MCNP6.2 code [9] with the base ENDF/B-VIII.0 nuclear data library [10]. The cross sections of activation reactions were taken from IRDFF-II library. Only aluminium block transport cross sections were modified using JEFF-3.3 [11], JENDL-5 [12], and ENDF/B-VIII.0 libraries. The neutron fluxes were calculated using the Next-Event Estimator, (neutron flux tally F5 in the MCNP6.2 code - point detector) with 40 groups per decade structure in the case of HPD and with steps of 100 keV in the case of the stilbene detector. The reaction rates were calculated using the neutron flux in the activation detector multiplied by the activation cross section of the dosimeter. The neutron flux was calculated using the averaged track-length estimator in a predefined region, divided by the volume of that region for one source particle per second (neutron flux tally F4 in MCNP6.2 code). The calculation model incorporated all available data into the MCNP model.

The aluminum has fine structures in its cross section which is reflected with fine structures in the neutron leakage flux. This fine structure is not fully reconstructed by measurement due to detector resolution. To reflect this effect, the calculations were broadened with regard to the determined resolution.

The input $^{252}$Cf(sf) reference neutron spectrum [13] was taken from the IRDFF-II webpage (https://www-nds.iaea.org/IRDFF/IRDFF-II_sp_ENDF.zip).

Results

Activation detectors

Table 3 shows graphical comparison of calculation (C) using various nuclear data libraries and expectation (E) in the form of a C/E-1 for Indium activation detectors monitoring $^{115}$In(n,n')$^{115m}$In reaction placed inside the aluminium block. Generally, the best agreement with the experiment is achieved with JENDL-5 library regardless of aluminium thickness. The best agreement is achieved for the thinnest thickness of the aluminium. For thicker thickness the agreement deteriorates. The ENDF/B-VIII.0 and JEFF-3.3 libraries gave the same results worser than JENDL-5.

Table 4 shows C/E-1 for the second set of activation foils placed inside the aluminium block. Generally, concerning threshold reactions, the best agreement is achieved with JENDL-5 library which gives slightly better results than other libraries. Worst results are achieved for the lowest thickness of aluminium, i.e. 1.18 cm; however results are loaded with high uncertainty. Concerning capture reactions, all libraries give similar results. Unsatisfactory results are achived for $^{197}$Au(n,γ)$^{198}$Au reaction regardless of thickness and library. $^{181}$Ta(n,γ)$^{182}$Ta fits into two sigma uncertainty for all libraries and thickness. $^{63}$Cu(n,γ)$^{64}$Cu agrees well for all libraries in the distance of 6.12 cm unlike the shortest distance.

*Table 3: C/E-1 comparison between calculated and expected reaction rates of $^{115}In(n,n')^{115m}In$ in various positions of aluminium block.*

| Al thickness [cm] | ENDF/B-VIII.0 [%] | JEFF-3.3 [%] | JENDL-5 [%] | Unc. [%] |
|---|---|---|---|---|
| 1.18 | -2.0 | -2.0 | -2.3 | 3.9 |
| 6.12 | -8.7 | -8.7 | -7.1 | 2.5 |
| 11.06 | -12.2 | -12.2 | -8.2 | 2.5 |
| 16.01 | -12.6 | -12.6 | -6.2 | 2.8 |
| 20.95 | -18.3 | -18.3 | -10.2 | 2.8 |
| 25.91 | -16.9 | -16.9 | -6.8 | 3.5 |

Fast neutron leakage spectrum

The C/E-1 plot for HPD detectors in the energy range of 0.1 MeV – 1.3 MeV is displayed in Figure 7. Overall, all libraries give reasonable results. ENDF/B-VIII.0 and JEFF-3.3 libraries give very similar results. ENDF/B-VIII.0 library underestimates experiment in the energy region of 0.75 -0.94 MeV and 1.06 – 1.3 MeV. JENDL-5 library overestimates experiment in the energy region of 0.94 -1.06 MeV and underestimates in the energy regions of 0.14 - 0.24 MeV and 0.75 - 0.89 MeV.

*Table 4: C/E-1 comparison between calculated and expected reaction rates of large foil set in various positions of the aluminum block.*

| Al thickness [cm] | Reaction | ENDF/B-VIII.0 [%] | JEFF-3.3 [%] | JENDL-5 [%] | unc. [%] |
|---|---|---|---|---|---|
| 1.18 | $^{197}Au(n,\gamma)^{198}Au$ | 14.7 | 14.5 | 14.4 | 4.3 |
| 1.18 | $^{63}Cu(n,\gamma)^{64}Cu$ | 18.3 | 18.3 | 18.0 | 8.4 |
| 1.18 | $^{46}Ti(n,p)^{46}Sc$ | -15.6 | -15.5 | -16.0 | 12.5 |
| 1.18 | $^{47}Ti(n,p)^{47}Sc$ | -14.0 | -13.9 | -14.1 | 11.7 |
| 1.18 | $^{58}Ni(n,p)^{58}Co$ | -9.4 | -9.4 | -9.6 | 12.2 |
| 1.18 | $^{181}Ta(n,\gamma)^{182}Ta$ | 0.5 | 0.7 | 0.6 | 8.1 |
| 1.18 | $^{92}Mo(n,p)^{92*}Nb$ | 11.4 | 11.4 | 10.9 | 12.4 |
| 6.12 | $^{197}Au(n,\gamma)^{198}Au$ | 18.4 | 18.5 | 18.5 | 2.5 |
| 6.12 | $^{63}Cu(n,\gamma)^{64}Cu$ | 0.5 | 0.4 | 0.0 | 4.1 |
| 6.12 | $^{46}Ti(n,p)^{46}Sc$ | 4.5 | 4.7 | 3.2 | 6.1 |
| 6.12 | $^{47}Ti(n,p)^{47}Sc$ | -2.1 | -2.0 | -1.1 | 4.2 |
| 6.12 | $^{58}Ni(n,p)^{58}Co$ | -7.2 | -7.2 | -6.5 | 4.3 |
| 6.12 | $^{181}Ta(n,\gamma)^{182}Ta$ | -3.4 | -3.5 | -3.0 | 3.3 |
| 11.06 | $^{197}Au(n,\gamma)^{198}Au$ | 24.9 | 25.2 | 25.3 | 2.4 |
| 11.06 | $^{58}Ni(n,p)^{58}Co$ | -8.7 | -8.4 | -6.6 | 2.9 |
| 11.06 | $^{181}Ta(n,\gamma)^{182}Ta$ | 2.2 | 2.3 | 3.3 | 3.3 |
| 16.01 | $^{197}Au(n,\gamma)^{198}Au$ | 32.7 | 32.5 | 33.7 | 2.7 |
| 16.01 | $^{58}Ni(n,p)^{58}Co$ | -11.9 | -11.5 | -10.0 | 2.6 |
| 16.01 | $^{181}Ta(n,\gamma)^{182}Ta$ | 3.7 | 4.0 | 5.3 | 3.0 |
| 20.95 | $^{58}Ni(n,p)^{58}Co$ | -8.0 | -8.0 | -2.7 | 2.4 |

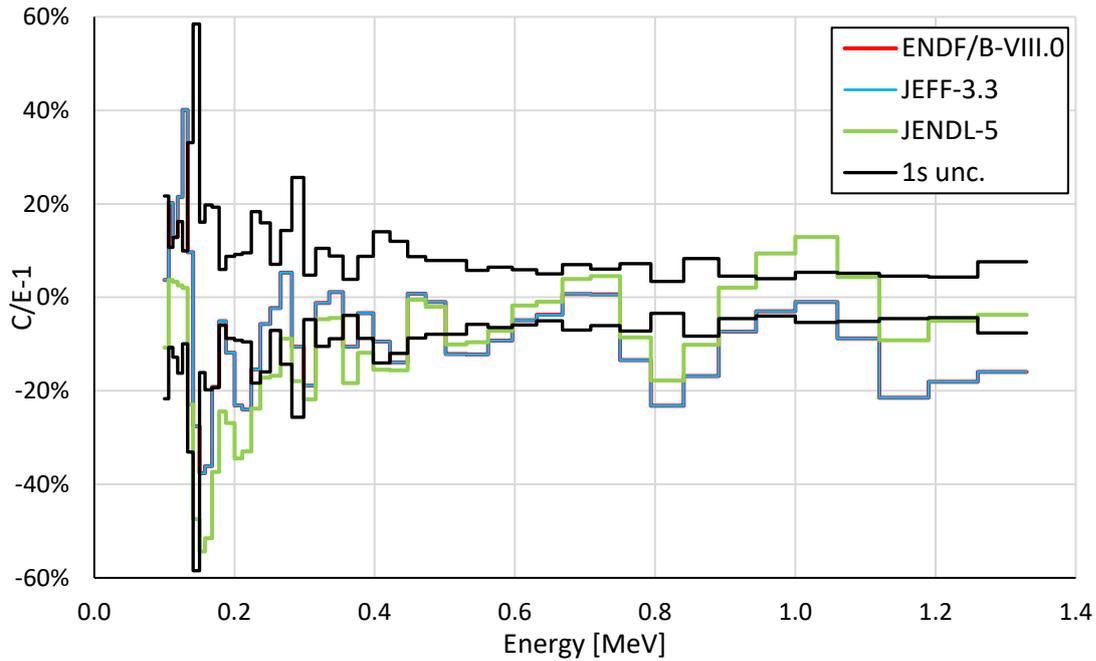

Figure 7: Comparison of measured neutron leakage flux with calculation using various nuclear data libraries with applied Gaussian broadening using experimentally determined resolution of HPD Detector.

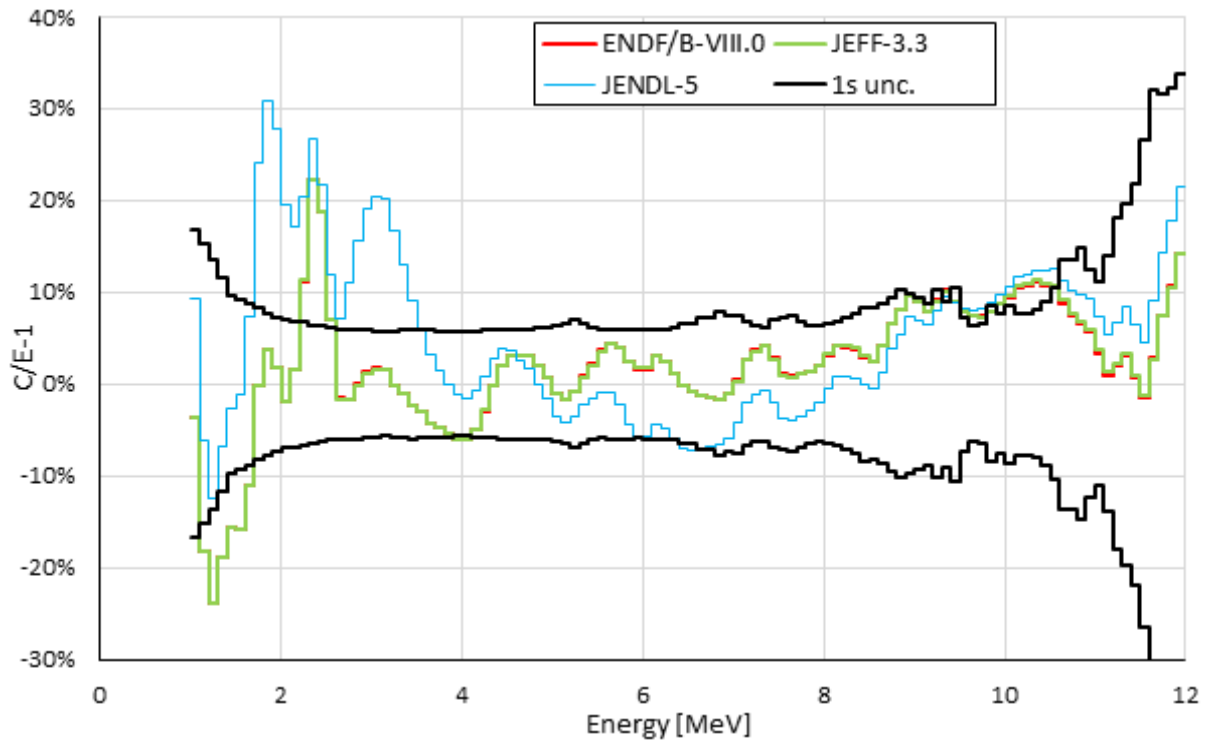

Figure 8: C/E-1 for different aluminium neutron transport libraries. Calculations are taken with applied Gaussian broadening. One sigma uncertainty is displayed as a bold black curve.

Figure 8 shows C/E-1 for stilbene scintillator measurement in the energy range of 0.9 MeV-13 MeV. The best agreement is achieved with ENDF/B-VIII.0 and JEFF-3.3 libraries. Their results are almost the same. The only region out of one sigma uncertainty falls into the approximate energy range of 2.3-2.6 MeV. JENDL-5 library disagrees with one sigma uncertainty in the energy region of 1.8 MeV-3.6 MeV.

Conclusions

In conclusion, the experimental validation of aluminium neutron transport libraries, conducted through various activation reactions and neutron flux measurements, provides valuable insights into the accuracy and reliability of different nuclear data libraries. The results indicate that all libraries generally offer good agreement with experimental data across most of the tested energy ranges, with ENDF/B-VIII.0 and JEFF-3.3 libraries demonstrating similar performance in fast neutron flux calculations. Furthermore, the activation reaction involving $^{197}$Au(n,γ)$^{198}$Au was notably less accurate regardless of the library or aluminum thickness. Overall, the study confirms the usefulness of these libraries in simulating neutron transport in aluminum, while also underscoring the need for further refinement in specific neutron energy ranges.

Acknowledgements

This study was co-funded/funded by European Union through the Euratom research and training programme EURATOM2027 under Grant Agreement No 101164596 in project APRENDE. We thank Jiří Pipster Malý for his hard work and mentoring.